\documentstyle[twocolumn,prb,aps,epsf]{revtex}

\input epsf
\begin{document}

\tightenlines

\twocolumn[\hsize\textwidth\columnwidth\hsize\csname @twocolumnfalse\endcsname

\title{Influence of finite Hund rules and charge transfer on properties of Haldane
systems.}
\author{A.E. Feiguin$^{a}$, Liliana Arrachea$^{b}$ and A.A. Aligia$^{c}$}
\address{$^{a}$Instituto de F\'{\i }sica Rosario (CONICET-UNR),\\
 Bv. 27 de febrero 210 bis, 2000 Rosario, Argentina\\
$^{b}$PUC-Rio, Departamento de Fisica,\\
Caixa Postal 38071, Cep: 22452-970 - RJ, Rio de Janeiro, Brasil\\
$^{c}$Centro At\'{o}mico Bariloche and Instituto Balseiro,\\
Comisi\'on Nacional de Energ\'{\i}a At\'{o}mica, \\
8400 Bariloche, Argentina.}
\maketitle
\vskip 1.cm

\begin{abstract}
We consider the Kondo-Hubbard model with ferromagnetic exchange coupling $%
J_{H}$, showing that it is an approximate effective model for late
transition metal-O linear systems. We study the dependence of the charge and
spin gaps $\Delta _{C}$, $\Delta _{S}$, and several spin-spin correlation
functions, including the hidden order parameter $Z(\pi )$, as functions of $%
J_{H}/t$ and $U/t$, by numerical diagonalization of finite systems. Except
for $Z(\pi )$, all properties converge slowly to the strong-coupling limit.
When $J_{H}/t\sim 2$ and $U/t\sim 7$ (the effective parameters that we
obtain for Y$_{2}$BaNiO$_{5}$), $\Delta _{S}$ is roughly half of the value
expected from a strong-coupling expansion.
\end{abstract}

\vskip .6cm]

\thispagestyle{empty}

\section{Introduction}

After Haldane's conjecture that integer-spin antiferromagnetic Heisenberg
chains should exhibit a gap in their excitation spectrum \cite{hal}, there
has been a considerable amount of research in spin $S=1$ systems \cite
{sev,aklt,ken}. Fascinating aspects of these systems are the presence of
free spin-1/2 excitations at the end of sufficiently long finite chains \cite
{aklt}, and the presence of a hidden string-topological order \cite
{ni,ken,gom}.

The compound Y$_{2}$BaNiO$_{5}$ is a candidate to a nearly ideal realization
of the spin $S=1$ antiferromagnetic Heisenberg chain and stimulated intense
research on the system recently. The 1D character is supported by
experimental evidence which shows that the exchange couplings transverse to
the chains are very small and unable to induce long range magnetic order 
\cite{xu}. The representation of the two $S=1/2$ holes per Ni by a $S=1$
 spin in the effective one-band model is, however, an intuitive but not
clearly justified simplification, being the underlying assumption a large
Hund rule acting on the two relevant Ni-orbitals. Such a point of view has
been adopted in most theoretical works of the system, like an alternative
interpretation of specific-heat experiments \cite{ram} in Zn-doped Y$_{2}$%
BaNiO$_{5}$ which raised doubts about the presence of free spin-1/2
excitations near the end of long chains \cite{dmrg}. Theoretical work
motivated by other experiments \cite{dit,koj} for Ca-doped Y$_{2}$BaNiO$_{5}$%
, for which the NiO chains are not broken, but doped with holes also retain
only the ground state of the 3d$^{8}$ configuration of Ni$^{+2}$ \cite
{lu,sor,pen,dag,epl}.

On general physical grounds one expects the effective Ni-Ni hopping (via O)
to be of the order of the corresponding one in CuO chains ($t\sim 0.85$eV 
\cite{vai}). Furthermore, spectroscopic data in atomic Ni show that the Hund
rule leads to a ferromagnetic exchange $J_{H}^{\prime }\sim 1.6$eV. Thus,
it seems that not only the triplet ground state of the 3d$^{8}$
configuration of Ni$^{+2}$, but also the excited singlet 
should be taken into account for a realistic description of the
system. Except for a brief discussion on the charge gap and Ni $L_{3}$ x-ray
absorption spectrum \cite{epl}, this issue has been unexplored so far, to
our knowledge. Even if the effective intratomic repulsion $U$ is large
compared with $t$, as we shall show, the properties of the system differ
from those of the strong-coupling limit. In particular, the expression $%
\Delta _{S}=0.41049J$ for the spin gap in terms of the effective Ni-Ni
exchange $J$ \cite{whi,gol}, is no longer valid (at least if $J$ is
calculated perturbatively, see section IV). Thus, at least a qualitative
study of the effects of a realistic $J_{H}$ (instead of infinite) on the
properties of these systems seems necessary.

In this work we derive and study the Kondo-Hubbard model with ferromagnetic
coupling, as an approximate (in the sense which will be clarified in the
next section) effective model for linear transition metal-oxygen systems, in
which the relevant transition-metal orbitals are the $e_{g}$ ones. The model
retains the effects of charge fluctuations at the transition-metal ions and
finite Hund rules. Explicit effective parameters are calculated for Y$_{2}$%
BaNiO$_{5}$. Its version with antiferromagnetic coupling $J_{H}$ has been
extensively studied in the context of heavy fermion systems \cite{review}
while the ferromagnetic case also adquired relevance in connection with the
physics of the perovskite Mn oxides with giant magnetoresistance \cite{penc}%
, being also closely related to the double-exchange model \cite{cie,zan} and
other models used to study these compounds \cite{ino,rie}. 

In section II, we explain the derivation of of the Kondo-Hubbard model as a
low-energy effective model, in which some terms of lower magnitude, were
neglected for simplicity. Section III contains the results for charge and
spin gaps, spin expectation values and several spin-spin correlation
functions, obtained by numerical diagonalization. Section IV contains the
conclusions.

\section{ The low-energy reduction procedure}

In the simplest and most usual perovskite structures, the transition metal
atoms are in sites of nearly cubic ($O_{h}$) symmetry, surrounded by six O
atoms, lying in the directions $\pm x$, $\pm y$, $\pm z$. In the particular
case of Y$_{2}$BaNiO$_{5}$, these octahedra are linked by their vertices and
form well separated chains, making a nearly ideal one-dimensional compound.
Since by far, the largest contribution to crystal-field splitting is due to
covalency \cite{sug}, near the end of the 3d series, the few holes present
in the 3d shell of the transition metal enter the $e_{g}$ \ (d$%
_{3z^{2}-r^{2}}$ or d$_{x^{2}-y^{2}}$) orbitals. This is due to their larger
hybridization with the 2p$_{\sigma }$ orbitals (those pointing towards the
transition metal atom) of the nearest-neighbor O atoms. The starting
multiband Hamiltonian for the system should include the above mentioned
orbitals and can be divided as follows:

\begin{equation}
H=H_{d}+H_{p}+H_{pd}+H_{pp},  \label{e2}
\end{equation}
where $H_{d}$ ($H_{p}$) describes the on-site correlations inside the 3d
(2p) shell. $H_{pd}$ contains the transition metal-O hoppings and
repulsions. The O-O hopping is described by $H_{pp}$. The last three terms
of Eq. (\ref{e2}) are a trivial extension of similar terms extensively
described and studied in multiband models for the cuprates \cite{h3b} and
will be not reproduced here. The largest energies in the problem and the
ingredients of the new physics when more than one relevant orbital per site
is present, are contained in $H_{d}$. The Coulomb and exchange integrals
among the $e_{g}$ spin-orbitals can be parameterized in terms of three
Slater parameters $F_{0}$, $F_{2}$ and $F_{4}$, using usual methods in
atomic physics \cite{con}. Denoting $a_{i\sigma }^{\dagger }$ ($b_{i\sigma
}^{\dagger }$) the creation operator for the $a_{1g}$ d$_{3z^{2}-r^{2}}$ ($%
b_{1g}$ d$_{x^{2}-y^{2}}$) orbital at site $i$ with spin $\sigma $, the
result can be written in the form:

\begin{eqnarray}
H_{d} &=&U_{d}\sum_{i}(a_{i\uparrow }^{\dagger }a_{i\uparrow }a_{i\downarrow
}^{\dagger }a_{i\downarrow }+b_{i\uparrow }^{\dagger }b_{i\uparrow
}b_{i\downarrow }^{\dagger }b_{i\downarrow })  \nonumber \\
&&+(U_{d}-J_{H}^{\prime })\sum_{i\sigma \sigma ^{\prime }}a_{i\sigma
}^{\dagger }a_{i\sigma }b_{i\sigma ^{\prime }}^{\dagger }b_{i\sigma ^{\prime
}}  \nonumber \\
&&+\frac{J_{H}^{\prime }}{2}\sum_{i}[\sum_{\sigma \sigma ^{\prime
}}a_{i\sigma }^{\dagger }b_{i\sigma ^{\prime }}^{\dagger }a_{i\sigma
^{\prime }}b_{i\sigma }  \nonumber \\
&&+(a_{i\uparrow }^{\dagger }a_{i\downarrow }^{\dagger }b_{i\downarrow
}b_{i\uparrow }+H.c.)],  \label{e3}
\end{eqnarray}
where $U_{d}=F_{0}+4F_{2}+36F_{4}$ and $J_{H}^{\prime }=8F_{2}+30F_{4}$.
Because of the neglect of the $t_{2g}$ (d$_{xy}$, d$_{yz}$, d$_{zx}$)
orbitals, $H_{d}$ lost the invariance under rotations of the atom, but
retains cubic ($O_{h}$) symmetry. From the two lowest excitation energies of
atomic Ni, we obtain $F_{2}=0.1600$eV and $F_{4}=0.0108$eV, leading to $%
J_{H}^{\prime }=1.60$eV. According to the theoretical interpretation of
optical experiments in NiO, $U_{d}+J_{H}^{\prime }/2=10$eV \cite{elp}. The
variation of $J_{H}^{\prime }$ along the 3d series is only a few per cent,
while the value of $U_{d}$ is very similar to that calculated in the
cuprates using constrained density-functional theory \cite{hyb}. More
sensitive to the particular system are the hopping parameters
(included in $H_{pp}$ and $H_{pd}$) and, particularly,
the transition-metal to O charge transfer energy $\Delta $ (defined as the
energy necesary to take a hole from the ground state 3d$^{8}$ configuration
of the transition metal and put it in the $p_{\sigma }$ orbital of the 2p
shell of a nearest-neighbor oxygen atom \cite{elp}), which increases to the
left of the periodic table.

To derive the effective Hamiltonian for one-dimensional transition metal-O
systems, we employ the cell perturbation method \cite{cell}. For simplicity
in the explanation below, we choose the particular case of NiO$_{6}$
octahedra sharing O atoms along the $z$ direction (present in Y$_{2}$BaNiO$%
_{5}$) and assume tetragonal symmetry. The $p_{z}$ orbitals of the O atoms
lying between two Ni atoms are expressed in terms of Wannier functions $\pi $
centered at each Ni atom \cite{note}. Each NiO$_{5}$ cell, composed of the
3d orbitals at one Ni site, the $\pi $ orbital at that site and those of the
four nearest O atoms along the $\pm x$ and $\pm y$ directions\cite{note3},
is solved exactly. To construct the low-energy effective Hamiltonian, only
eight eigenstates of the cell are retained: for two holes, the $B_{1g}$
triplet (which is essentially the ground-state of the Ni$^{+2}$
configuration plus corrections due to hybridization), and the first excited
state, the $B_{1g}$ singlet. For one and two holes in the cell, the lowest $%
B_{1g}$ doublets are retained. These eight states are mapped into the
corresponding ones of the Kondo-Hubbard model $H_{KH}$ (Eq. \ref{e4}) at the
corresponding site. The matrix elements of $H$ in the restricted basis are
calculated and mapped into the corresponding ones of $H_{KH}$. The effect of
the remaining states of $H$ could be included perturbatively but we neglect
it for simplicity. To retain a simple and more general form of $H_{KH}$, we
also neglect the dependence of the resulting effective hopping on the
occupation and spin of the sites involved. The resulting effective
Hamiltonian has the form:

\begin{eqnarray}
H_{KH} &=&-t\sum_{i}a_{i\sigma }^{\dagger }a_{i+1\sigma
}+U\sum_{i}a_{i\uparrow }^{\dagger }a_{i\uparrow }a_{i\downarrow }^{\dagger
}a_{i\downarrow }  \nonumber \\
&&-J_{H}\sum_{i}{\bf S}_{ia}\cdot {\bf S}_{ib},  \label{e4}
\end{eqnarray}
where ${\bf S}_{ia}$, ${\bf S}_{ib}$ are the spin of the fermions
represented by the hole creation operators $a_{i\sigma }^{\dagger }$ and $%
b_{i\sigma }^{\dagger }$ respectively. These are effective operators with
the same symmetry as those entering Eq. (\ref{e3}), but which differ from
them in the general case. In what follows $a_{i\sigma }^{\dagger }$ and $%
b_{i\sigma }^{\dagger }$ refer to these effective operators.

The meaning of $H_{KH}$ is easier to understand in the limit $\Delta \gg
U_{d}\gg J_{H}^{\prime }$, $t_{pd}$, where $t_{pd}$ is the Ni-O hopping
along the chain. In this case, the interactions $U$ and $J$ of the effective
Hamiltonian (\ref{e4}) coincide with those of $H_{KH}$, i.e. $U\sim U_{d}$
and $J_{H}\sim J_{H}^{\prime }$, while the effective hopping in (\ref{e4})
is given by the second-order process which carries a 3d$_{3z^{2}-r^{2}}$
hole to the same orbital of a nearest-neighbor Ni atom: $t=t_{pd}^{2}/\Delta 
$. However, the case  $\Delta \gg U_{d}$ is not representative of charge
transfer systems like Y$_{2}$BaNiO$_{5}$, for which $\Delta <U_{d}$. In this
case, the states $a_{i\uparrow }^{\dagger }a_{i\downarrow }^{\dagger
}b_{i\sigma }^{\dagger }|0\rangle $ actually represent states with
occupation close to one in the O $\pi $ orbitals.  

As a consequence, $U$ is mainly determined by $\Delta $ instead of $U_{d}$,
and the hopping matrix elements become dependent on the occupation and spin
of the two sites involved \cite{note2}. As mentioned above, this dependence
was neglected to keep a simple and more general form of $H_{KH}$.

To estimate the parameters of the effective model for Y$_{2}$BaNiO$_{5}$, we
took the values of $J_{H}^{\prime }$ and $U_{d}$ mentioned above, and the
(more uncertain) values of $\Delta $ and the different hopping parameters
 in $H_{pd}$ and $H_{pp}$ were taken from work on NiO \cite{elp},
with the $p-d$ and $p-p$ hopping parameters scaled with distance $r$ as $%
r^{-7/2}$ and $r^{-2}$ respectively. The parameters of $H_{KH}$ which result
from the mapping procedure are $U=4.4 $eV, $J_{H}=1.2$eV and $t\sim 0.7$eV.
It is interesting to note that $J_{H}$ has a very small sensitivity to the
parameters of $H$. Instead, changing $\Delta $ and the hoppings of $H$
within reasonable values affects $U$ by $\sim 20\%$ and $t$ by $\sim 30\%$.

\section{Results.}

In this section, we study the behavior of the charge and spin gap, spin
expectation values and spin-spin correlation functions of $H_{KH}$, using
Lanczos diagonalizations in periodic rings of length $L=4$, 6 and 8. The
rapid increase of the Hilbert space with $L$ prevents us to study longer
even chains with the present state of the art, but as we shall show, some
trends are already clear. The unit of energy is chosen as $t=1$.

\subsection{Charge gap}

In Fig. 1 we represent the charge gap $\Delta _{C}=E(1)+E(-1)-2E(0)$, where $%
E(n)$ is the ground-state energy for $n$ added holes to the stoichiometric
system (which contains one $a_{1g}$ and one $b_{1g}$ hole per site). The
result for $L=8$ is compared to that of a polynomial extrapolation in $1/L$
to estimate finite-size effects. These effects are small for $U\ \geq 4$. As
expected, the gap increases with $U$ and $J_{H}$. In the strong-coupling
limit $t=0$, the gap is $\Delta _{0}=U+$ $J_{H}/2$. As $t$ is turned on, but
kept small, the leading correction to $E(0)$ is of order $t^{2}/\Delta _{0}$%
, while those of $E(1)$, $E(-1)$ are equal and of order $t$. Assuming a Neel
background (alternating spin projections 1 and -1) the correction for one
added or one removed hole can be calculated and is $-\sqrt{2}t$. In both
cases, it is more convenient to align ferromagnetically the spin at the site
of the added or removed hole with those of its nearest neighbors. Thus, for
large $U$, $J_{H}$, we estimate:

\begin{equation}
\Delta _{C}=\Delta _{0}-2\sqrt{2}t.  \label{e5}
\end{equation}

In Fig. 2 we represent $\Delta _{C}-\Delta _{0}$ as a function of $U$ and $%
J_{H}$. The results agree with Eq. (\ref{e5}) in the strong-coupling limit.
In the opposite limit, for $U=0$, and small values of $J_{H}$, the results
have important size effects, and a large positive value of $\Delta
_{C}-\Delta _{0}$ is not reasonable. However, the extrapolated results show
a reasonable behavior and tend to small values in the limit of $J_{H}=0$. In
any case, the results for $U\geq 4$ seem reliable. From the parameter
estimates for Y$_{2}$BaNiO$_{5}$ given at the end of the previous section ($%
U/t\cong 6.3$, $J_{H}/t\cong 1.7$, $t\cong 0.7$eV), we obtain $\Delta
_{C}\sim 3$eV. This is somewhat larger than the experimental value $\Delta
_{C}\sim 2$eV \cite{dit}. This discrepancy is probably due to the fact that
the charge transfer gap and possibly the  hopping parameters cannot be
transferred directly from Ref. \cite{elp} (which is a theoretical
interpretation of optical spectra in NiO) to Y$_{2}$BaNiO$_{5}$.

\subsection{Spin gap}

In Fig. 3, we show the spin gap $\Delta _{S}$ as a function of $J_{H}$ for
several values of $U$. For $U=0$, the result for $\Delta _{S}$ has already
been reported \cite{tsun}. Here, for the smaller values of $U$ and $J_{H}$,
the finite-size effects are too large, and the extrapolated values to the
thermodynamic limit (in some cases negative) are meaningless. However, for
more realistic values of $U$, our results allow us to extract some
conclusions.

The qualitative behavior of $\Delta _{S}$ as a function of $J_{H}$ was to be
expected from the limiting cases: if $J_{H}=0$, the model is equivalent to
the Hubbard model plus $L$ free spin-1/2 states, and therefore $\Delta
_{S}=0 $. For $J_{H}\rightarrow 0$, $\Delta _{S}\sim $ $J_{H}^{2}$ has been
obtained using bosonization \cite{fuji}. In the limit of large $J_{H}$, the
low-energy physics of $H_{KH}$ reduces to a spin-1 Heisenberg chain:

\begin{equation}
H_{Heis}=J\sum_{i}{\bf S}_{i}\cdot {\bf S}_{i+1},\;\;\;\;\text{ }J=\frac{%
t^{2}}{\Delta _{0}}=\frac{t^{2}}{U+J_{H}/2},  \label{e6}
\end{equation}
where ${\bf S}_{i}={\bf S}_{ia}+{\bf S}_{ib}$. This model has a spin gap $%
\Delta _{S}^{Heis}=c(L)J$, where the constant $c(L)$ depends on the size of
the system. In particular $c(4)=1$, $c(6)=0.72$, $c(8)=0.59$, $c(\infty
)=0.41049$ \cite{gol}. The first three values of $c(L)$ coincide within 1\%
with our results extrapolated to infinite $J_{H}$. As $J_{H}$ increases,
first $\Delta _{S}$ increases from zero, and as the system approaches the
strong-coupling limit, $\Delta _{S}$ decreases with the effective spin-1
exchange $J$.

In Fig. 4 we show $\Delta _{S}/\Delta _{S}^{Heis}$ as a function of $J_{H}$.
For $U>4$, the extrapolated values do not differ very much from those of $%
L=8 $. Note that even for large values of $U$, the ratio $\Delta _{S}/\Delta
_{S}^{Heis}$ is considerably smaller than 1 if $J_{H}\sim t$. In particular,
for the parameters estimated for Y$_{2}$BaNiO$_{5}$, $\Delta _{S}/\Delta
_{S}^{Heis}$ $\sim 0.5$. However, this ratio was assumed 1 to estimate the
value of $J$ from experimental measurements of $\Delta _{S}$. Using the same
set of parameters, we obtain from the extrapolated values $\Delta _{S}\sim
240$K, while the experimental value is $\Delta _{S}\sim 100$K \cite{xu,dit}.
In view of the approximations made in deriving $H_{KH}$, the uncertainties
in the parameters of $H$, and the sensitivity of $\Delta _{S}$ and $J$ to
these parameters, the result is satisfactory. Part of the overestimate is
due to ferromagnetic corrections to $J$ in second-order in the intercell
hopping, which involve virtual quadruplet three-hole states. These states
are contained in $H$, but were projected out of the Hilbert space of $H_{KH}$%
. We have calculated $J$ by the cell-perturbation method, including these
corrections. The effective $J$ is reduced from 0.098eV to 0.088eV, but the
result is quite sensitive to the parameters of $H$. Comparison with exact
diagonalizations of a Ni$_{2}$O$_{11}$cluster \cite{epl}, shows that the
second-order result of the cell-perturbation method is still an
overestimation by a factor near 2, due to higher order corrections.

\subsection{Spin expectation values}

The behavior of the spin gap as a function of $J_{H}$ and $U$ displays a
slow change from the weak to the strong-coupling regimes. In Fig. 5(a) we
show the ratio of the spin gap to the effective exchange, as a function of $%
U $ for $L=8$ and different values of $J_{H}$. For any non-zero value of $%
J_{H} $, the strong-coupling limit is reached for sufficiently large $U$ and
the gap tends to the limit $\Delta _{S}^{Heis}=0.59J$. Fig 5(b) shows the
corresponding change in the total spin of both itinerant ($a_{1g}$) and
localized ($b_{1g}$) holes as $U$ is increased, in the lowest-energy state
with total spin and projection $S_{t}=S_{t}^{z}=1$. In the limit of small $%
J_{H}$, both types of holes are decoupled and it is easier to flip a
localized hole rather than an itinerant one from the $S_{t}=0$ ground state.
As a consequence, $S_{bt}^{z}\cong 1$, $S_{at}^{z}=1-S_{bt}^{z}\cong 0$. In
the opposite limit of very large $J_{H}$, the singlet states at each site ($%
(a_{i\uparrow }^{\dagger }b_{i\downarrow }^{\dagger }-a_{i\downarrow
}^{\dagger }b_{i\downarrow }^{\dagger })|0\rangle $) can be projected out of
the relevant Hilbert space, and in this case, the following equality among
spin operators at a given site can be proved: $2{\bf S}_{ia}=2{\bf S}_{ib}=%
{\bf S}_{i}$. Summing over all sites: $2{\bf S}_{at}=2{\bf S}_{bt}={\bf S}%
_{t}$. Thus, $S_{at}^{z}$ changes from 0 to 1/2 as $J_{H}$ increases and $%
S_{bt}^{z}=1-S_{at}^{z}$. The effect of increasing $U$ is to localize the
itinerant $a_{1g}$ holes and therefore, to contribute to the effect of $%
J_{H} $, reaching faster the strong-coupling limit. However, it is
noticeable that the approach to this limit is very slow. Comparison of the
quantities represented in Fig. 5 for different sizes suggests that this
approach is even slower in the thermodynamic limit.

\subsection{Spin-spin correlations}

In addition to the spin-spin correlation functions 
\begin{eqnarray}
S_{1}(l) &=&\langle
(S_{ia}^{z}+S_{ib}^{z})(S_{i+la}^{z}+S_{i+lb}^{z})\rangle ,  \nonumber \\
S_{2}(l) &=&\langle
(S_{ia}^{z}-S_{ib}^{z})(S_{i+la}^{z}-S_{i+lb}^{z})\rangle ,  \label{e7}
\end{eqnarray}
we also study in this section the string correlation function 
\begin{equation}
Z(j-i)=\langle S_{i}^{z}\exp (i\pi
\sum_{l=i+1}^{j-1}S_{l}^{z})S_{j}^{z}\rangle .  \label{e1}
\end{equation}
This latter has been propposed as a hidden order parameter for $S=1$ chains
to describe a hidden Z$_{2}\times $ Z$_{2}$ symmetry breaking corresponding
to the appearance of the Haldane gap \cite{ni,ken}. This symmetry has been
first implicitly introduced in an elegant variational approach for the
excited states \cite{gom}. In Fig. 6, we show the different correlation
functions. $S_{2}(l)$ has an on-site value $S_{2}(0)\cong 0.3$ and for other
distances $S_{2}(l)<0.02$ for the parameters of Fig. 6. The
antiferromagnetic correlations are evident. They are larger for the
localized holes than for the itinerant ones, as expected. For $U=0$, a
tenedency to antiferromagnetic order with wave vector $q=\pi $ is expected
from the form $\cos (2k_{F}r)$ of the oscilating
Ruderman-Kittel-Kasuya-Yosida effective interaction between localized $%
b_{1g} $ holes at a distance $r$ mediated by the mobile $a_{1g}$ ones with
Fermi wave vector $k_{F}=\pi /2$. In the strong-coupling limit, the
effective model $H_{Heis\text{ }}$(\ref{e6}) leads to the same type of
short-range correlations.

The Fourier transform ($S_{1}(q)=\sum_{l}e^{-iql}S_{1}(l)$, etc.) of some of
these correlation functions for wave vector $q=\pi $, is represented in Fig.
7. As $U$ increases, $S_{1}(\pi )$ and $S_{2}(\pi )$ approach slowly the
asymptotic value in the strong-coupling limit, as it was the case of the
spin gap and spin expectation values already discussed. Instead, $Z(\pi )$
seems to saturate faster to a fixed value as $J_{H}$ and $U$ increase.

In the strong-coupling limit, $Z(\pi )$ is a signature of the Haldane state 
\cite{ni,ken}. For our model, $Z(\pi )$ is a possible generalization of this
order parameter, when local singlet states and charge fluctuations are
allowed. The results of Fig. 7 are a hint that $Z(\pi )$ can be used as the
corresponding parameter of a hidden order that also exists in the
Kondo-Hubbard model $H_{KH}$. The difference in the behavior of $Z(\pi )$ as
a funcion of $U$, $J_{H}$, in comparison with that of the other correlation
functions, is an indication that $Z(\pi )$ is more sensitive to the
transition to the spin-gap state. That is precisely what one expects from a
quantity playing the role of an order parameter. The other spin-spin
correlations should be short-range-like, 
due to the opening of the spin gap for finite $U, J_{H}$ \cite{tsu,fuji}.
 Bosonization
results indicate that this gap increases quadratically with the Hund
coupling $J_{H}$ \cite{fuji}, and is thus very small for $J_{H}\rightarrow 0$%
. In this regime, the correlation length is larger than the maximum system
size that we have studied and we are unable to identify the change in the
behavior of the correlation functions. For larger values of $J_{H}$, the
change of regime is captured by our results, and more clearly from the
behavior of $Z(\pi )$.

\section{Summary and discussion}

We have studied charge and spin gap, spin expectation values, and several
spin-spin correlation functions of a Kondo-Heisenberg model $H_{KH}$, for
two particles per site. We have shown that the model can be considered as an
approximate effective model for one-dimensional transition metal-O systems,
in which only the $e_{g}$ orbitals of the transition metals are relevant.
Without any adjustable parameters (taking the parameters of the original
multiband Hamiltonian from NiO scaled appropriately with distance), we
obtain from the effective $H_{KH}$, a charge and a spin gap of the correct
order of magnitude for Y$_{2}$BaNiO$_{5}$. The model has also been used as a
simplified model for the manganites, and our results should be qualitatively
valid in the limit in which all Mn ions are Mn$^{+3}$.

For sufficiently large $U$ or $J_{H}$, the charge gap of $H_{KH}$ is
approximately given by $\Delta _{C}\cong U+J_{H}/2-2\sqrt{2}t$. The
effective model $H_{KH}$ contains the spin-1 antiferromagnetic Heisenberg
model (also called Haldane chain) in the strong-coupling limit $t\ll $ $%
J_{H},U$. We obtain however that for realistic parameters for Y$_{2}$BaNiO$%
_{5}$ or when the effective Hund-rule exchange coupling $J_{H}$ is not
much larger than the effective hopping $t$, several properties differ from
the Haldane limit. In particular, the different dynamics of the itinerant
and mobile $e_{g}$ holes (reflecting that they do not behave as part of the
same spin-1 object) are clearly manifested in spin-spin correlation
functions. In addition, the spin gap $\Delta _{S}$ is roughly half of that
expected from a strong coupling expansions. This fact should be taken into
account when the effective spin-1 exchange $J$ is extracted from
experimental values of $\Delta _{S}$ and in the consistent interpretation of
different thermodynamic experiments together with $\Delta _{S}$. One
possible way to interpret our results for $\Delta _{S}$ when $U\gg J_{H}\sim
t$, is that the effective spin-1 Heisenberg Hamiltonian Eq. (\ref{e6}) is
still valid, but higher order corrections in $t$ reduce appreciably the
second-order result $J=t^{2}/(U+J_{H}/2)$ for the effective exchange.
Fourth-order corrections which include local singlet states as intermediate
states are consistent with this reduction. However, when $t\sim J_{H}$,
perturbation theory ceases to be valid, and it seems more adequate to
include the local singlets ($(a_{i\uparrow }^{\dagger }b_{i\downarrow
}^{\dagger }-a_{i\downarrow }^{\dagger }b_{i\downarrow }^{\dagger
})|0\rangle $ in our notation) in the model Hamiltonian.

\section*{Acknowledgments}

Two of us are supported by the Consejo Nacional de Investigaciones
Cient\'{\i}ficas y T\'{e}cnicas (CONICET), Argentina. A. A. A. is partialy
supported by CONICET. We would like to thank to H.A. Ceccatto for useful
discussions.

{\large {\bf FIGURE CAPTIONS}}

{\bf Fig. 1.} Charge gap as a function of $J_{H}$ for several values of $U$
indicated inside the figure. Solid symbols are the result for $L=8$ sites.
Open symbols correspond to extrapolation to the thermodynamic limit from the
results of $L=4,6$ and 8, using a quadratic polynomial in $1/L$.

{\bf Fig. 2. }Same as in Fig. 1 for the charge gap minus its strong-coupling
value $\Delta _{0}=U+J_{H}/2$.

{\bf Fig. 3.} Spin gap as a function of $J_{H}$ for several values of $U$.
Solid symbols correspond to $L=8$ and open symbols to the extrapolated value.

{\bf Fig. 4. }Ratio of the spin gap $\Delta _{S}$ over its strong-coupling
value ($\Delta _{S}^{Heis}=c(L)J$, where $J=t^{2}/(U+J_{H}/2)$, see text) as
a function of $J_{H}$. The meaning of the symbols is the same as before.

{\bf Fig. 5. }(a) Ratio of the spin gap to the effective exchange $\Delta
_{S}/J$ as a function of $U$ for $L=8$ and several values of $J_{H}$: $%
J_{H}= $ 2 (circles), $J_{H}=$ 4 (squares), $J_{H}=$ 6 (diamonds) and $%
J_{H}= $ 20 (dashed line) . (b) $z$ component of the total spin of the $%
a_{1g}$ ($b_{1g}$) holes as a as a function of $U$ for $L=8$ and several
values of $J_{H}$ indicated by the same solid (open) symbols as above, in
the lowest-energy state with total spin and $z$ component $S_{t}=S_{t}^{z}=1$%
.

{\bf Fig. 6. }Spin-spin correlation functions $\langle
S_{i}^{z}S_{i+l}^{z}\rangle $ as a function of distance $l$ for $a_{1g}$ ($%
S_{a}(l)$) holes, $b_{1g}$ ($S_{b}(l)$) holes, the sum of both spins ($%
S_{1}(l)$), and that defined by Eq. (1), for $J_{H}=$ 2 and two values of $%
U:U=$ 2 (circles), and $U=10$ (diamonds).

{\bf Fig. 7. }Fourier transform of the correlation functions at momentum $%
\pi $ for the sum ($S_{1}(\pi )$) and difference ($S_{2}(\pi )$) of the spin
of both types of holes at a given site, and the hidden order parameter $%
Z(\pi )$.

{\bf Fig. 8. } Size dependence of $S_{1}(q)$ and $Z(q)$ for $J_{H}=2$ , $%
U=10 $.

\newpage
\leavevmode

\begin{figure}
\begin{center}
\epsfbox{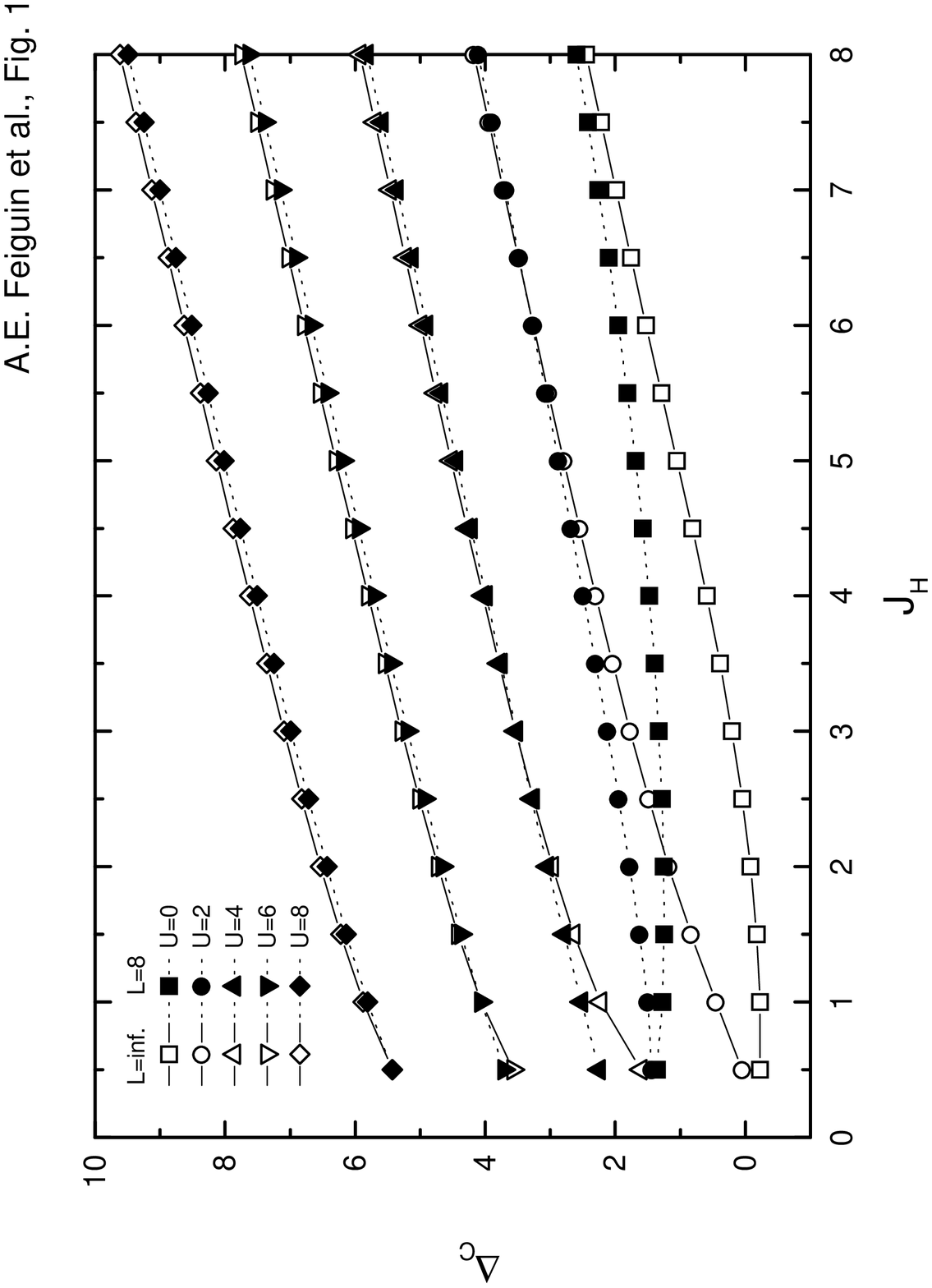}
\end{center}
\end{figure}

\begin{figure}
\begin{center}
\epsfbox{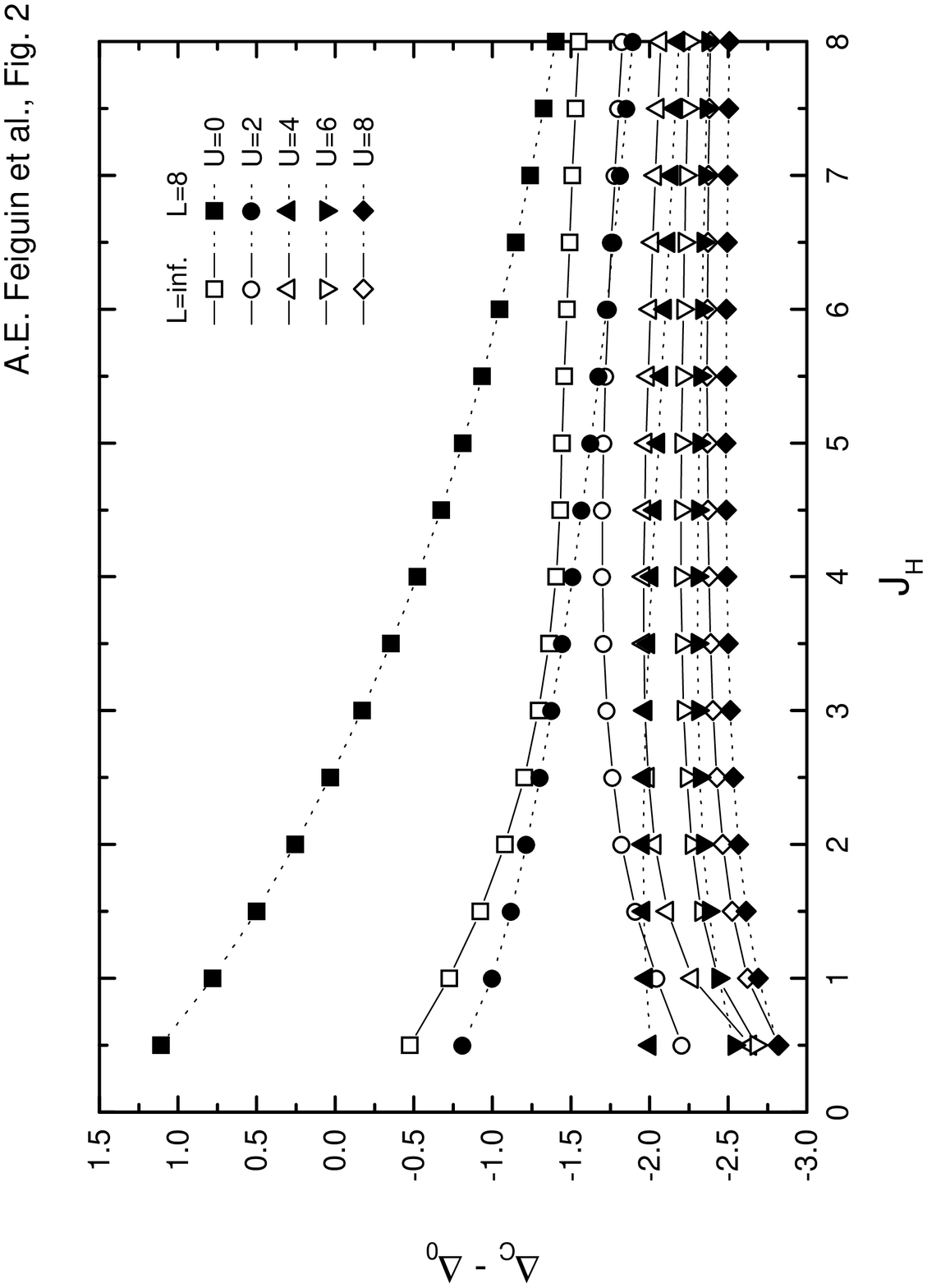}
\end{center}
\end{figure}

\begin{figure}
\begin{center}
\epsfbox{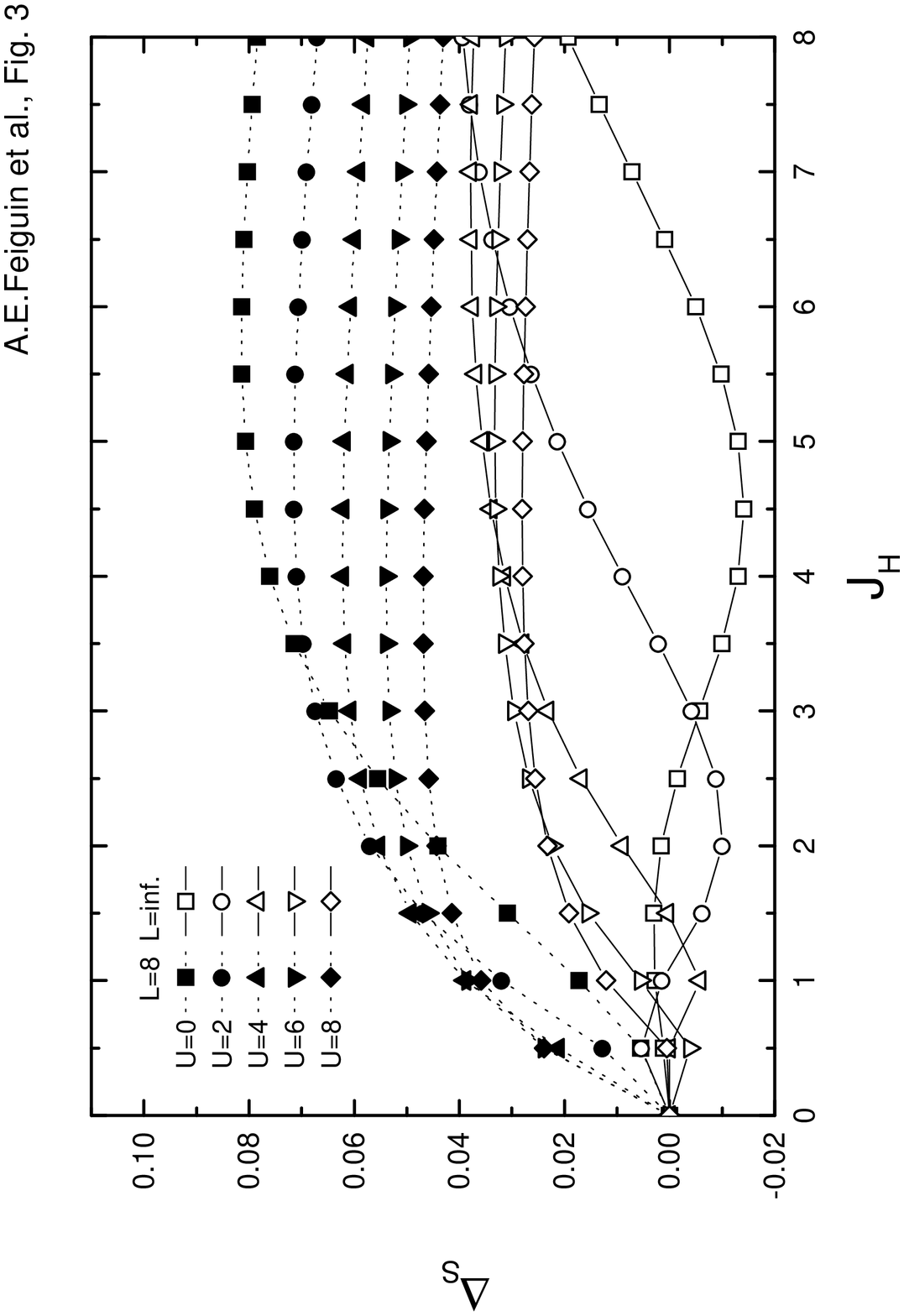}
\end{center}
\end{figure}

\begin{figure}
\begin{center}
\epsfbox{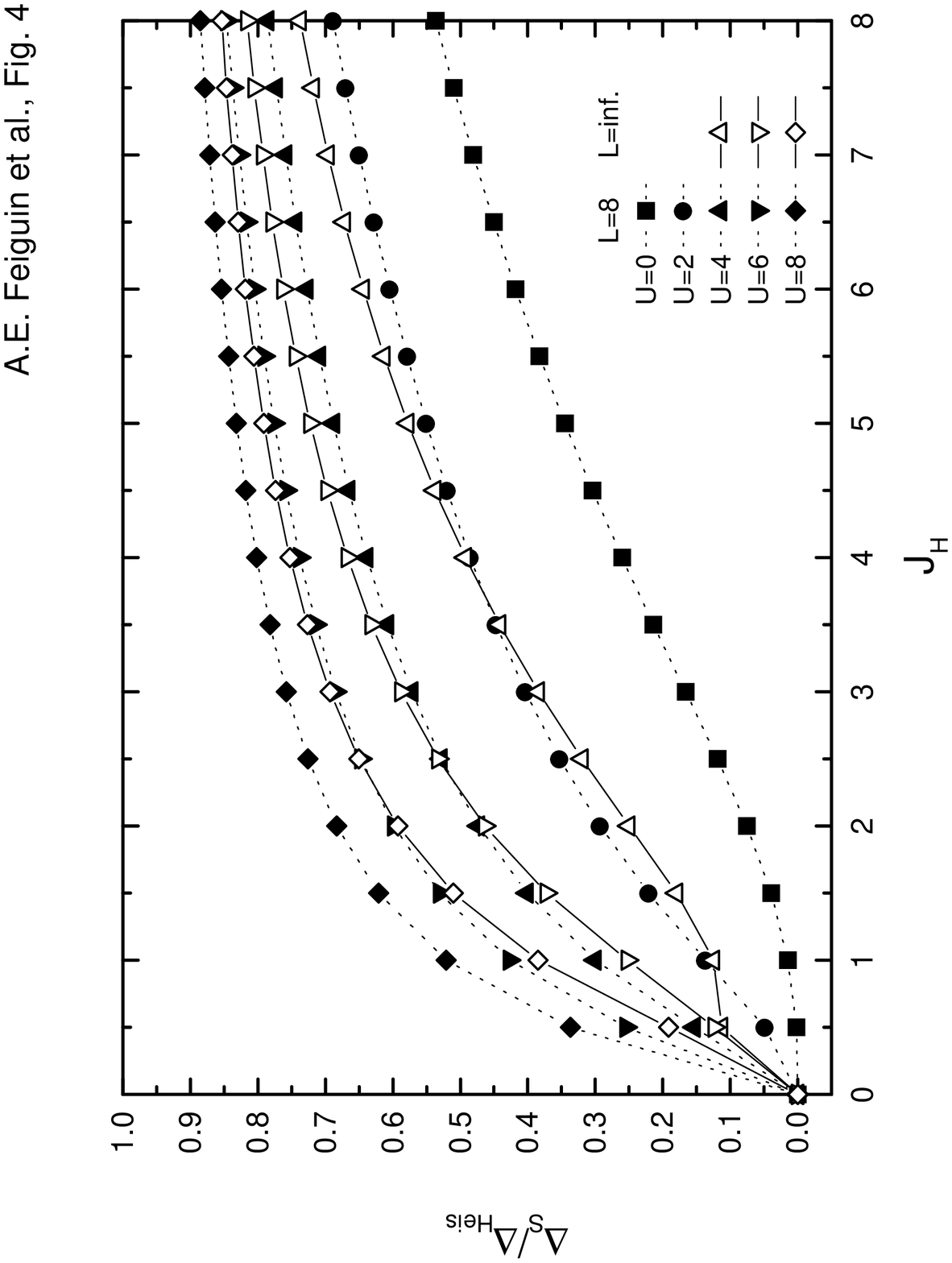}
\end{center}
\end{figure}

\onecolumn
\samepage

\begin{figure}
\begin{center}
\epsfbox{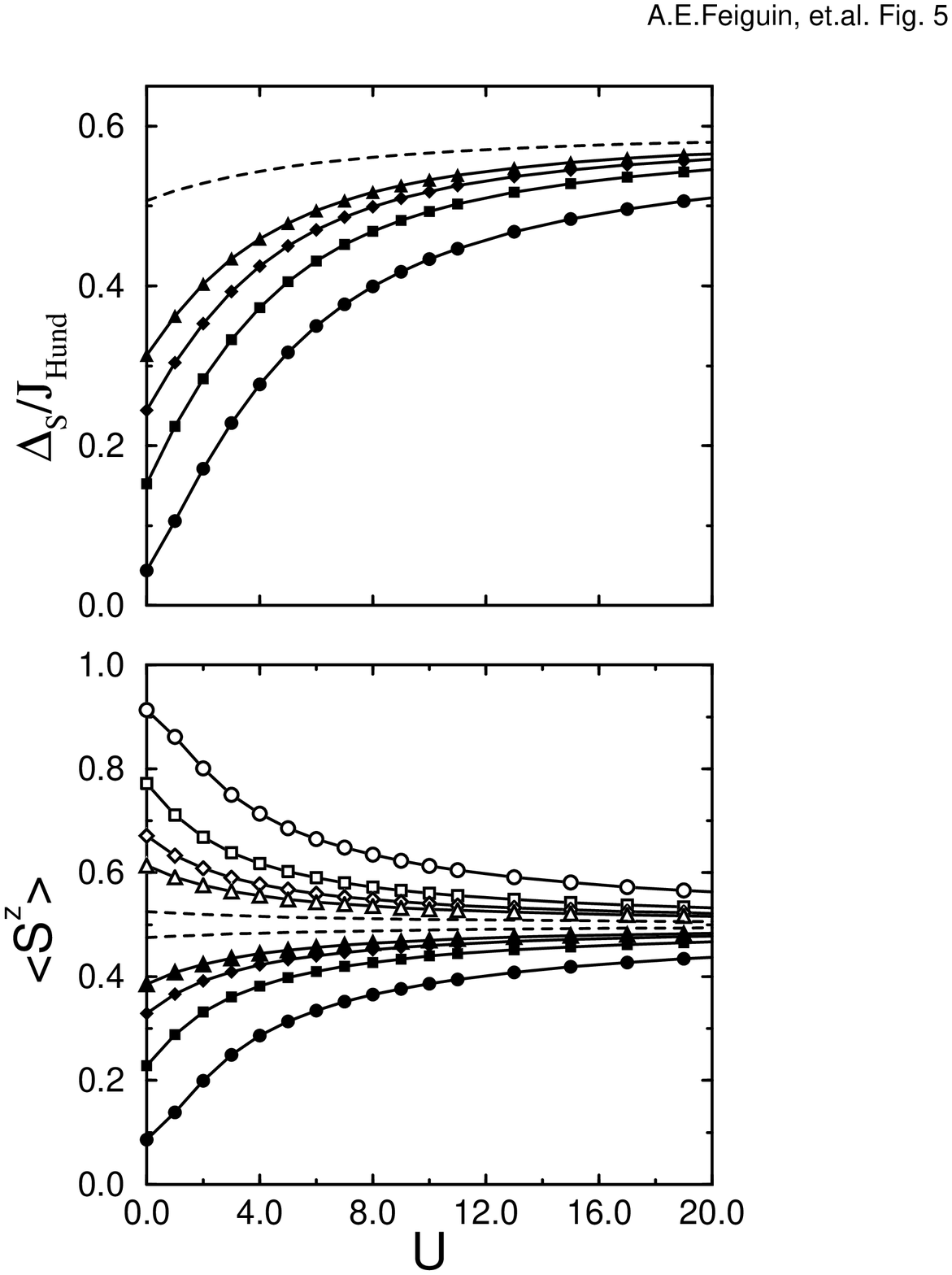}
\end{center}
\end{figure}

\newpage

\begin{figure}
\begin{center}
\epsfbox{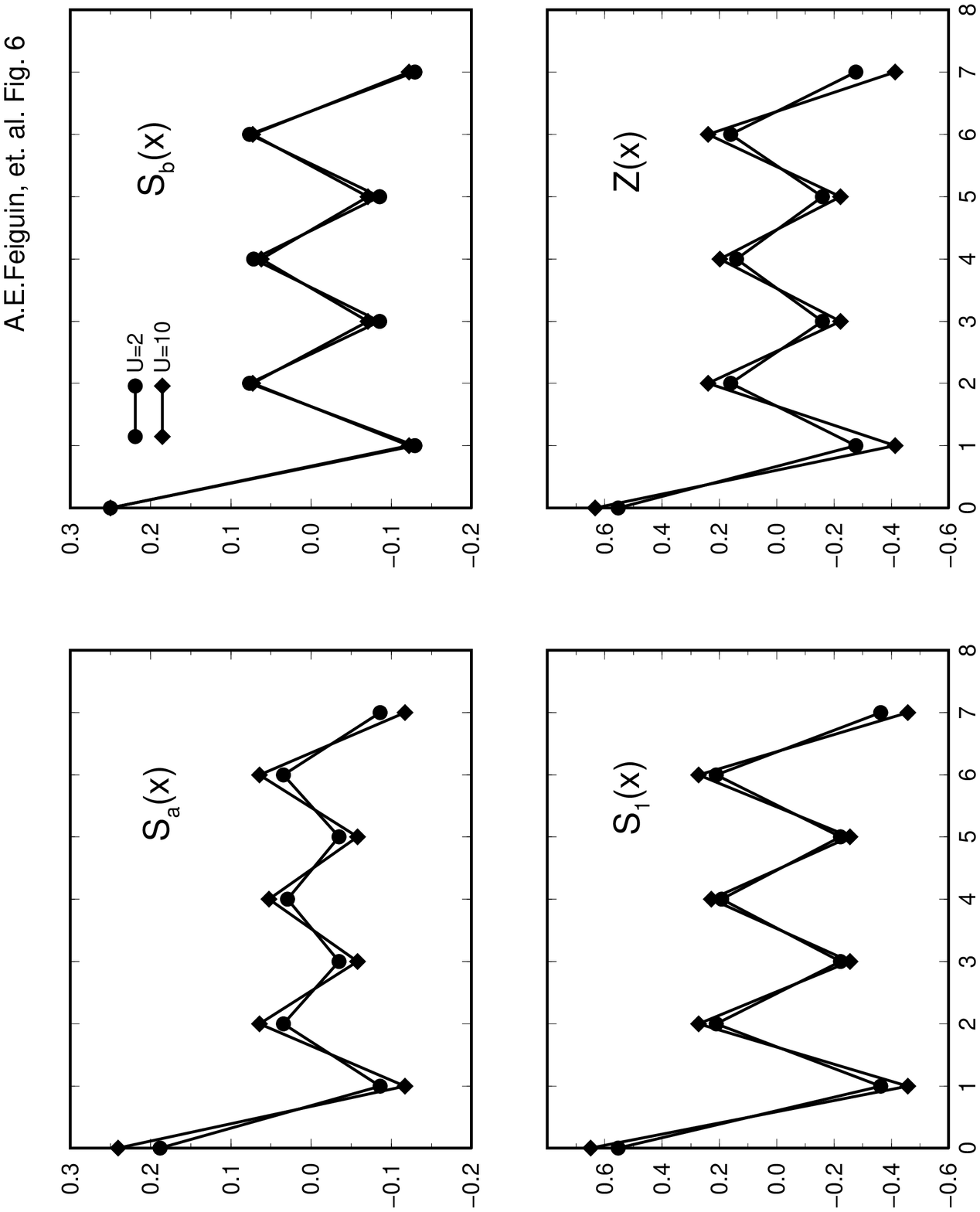}
\end{center}
\end{figure}

\newpage

\begin{figure}
\begin{center}
\epsfbox{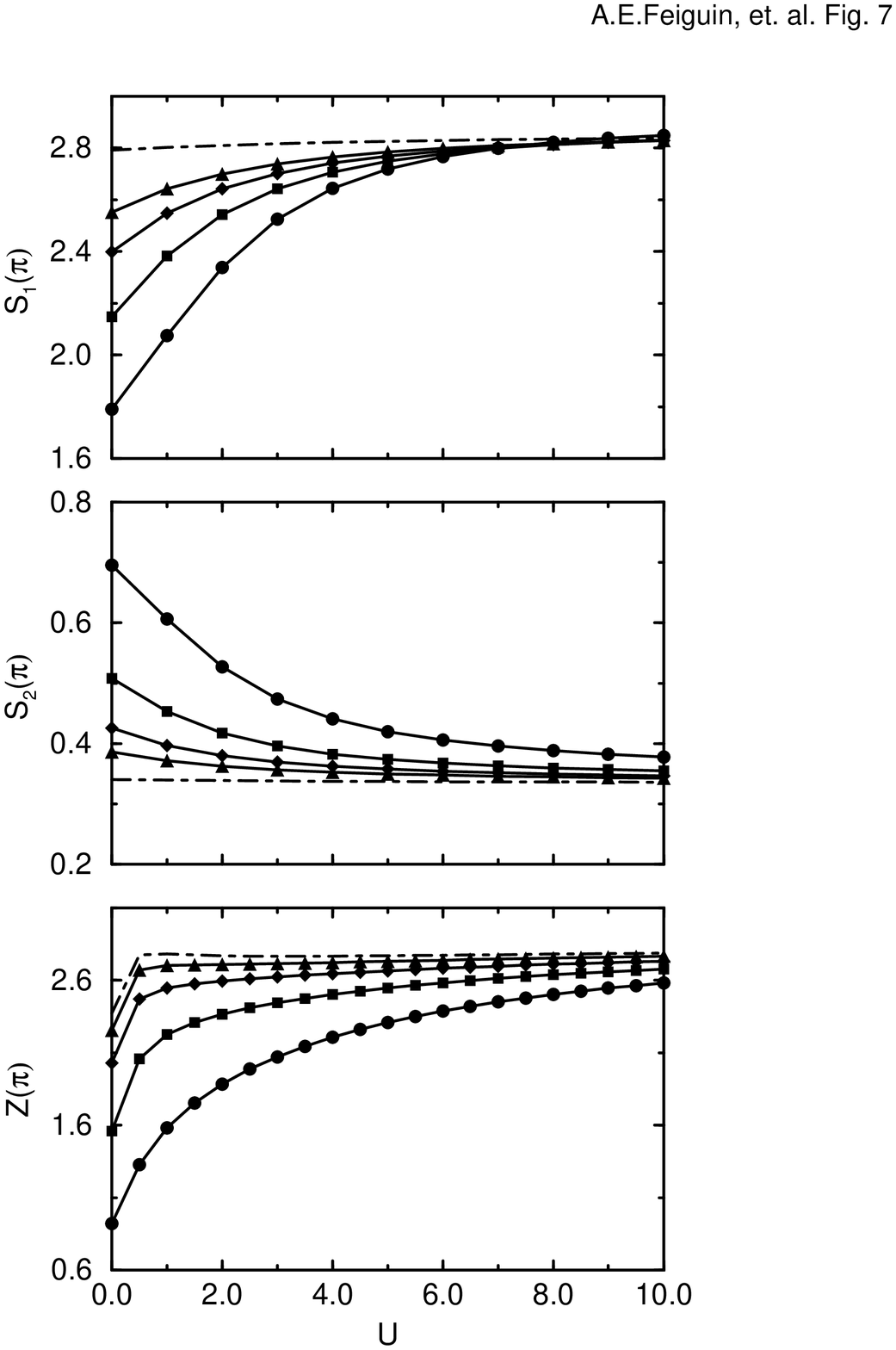}
\end{center}
\end{figure}

\newpage

\begin{figure}
\begin{center}
\epsfbox{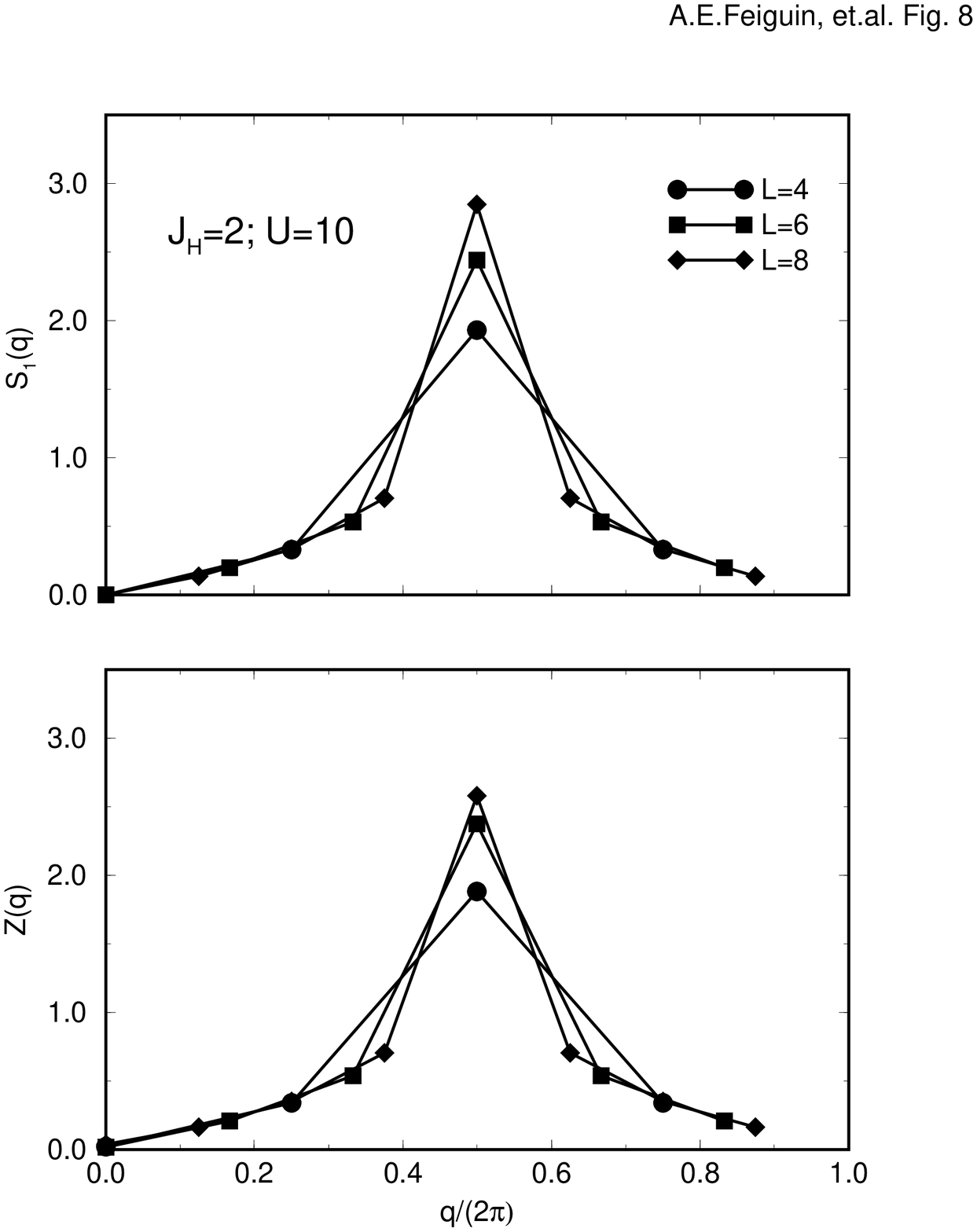}
\end{center}
\end{figure}


\begin{references}
\bibitem{hal}  F.D.M. Haldane, Phys. Rev. Lett. {\bf 50}, 1153 (1983); Phys.
Lett. {\bf 93A}, 464 (1983).

\bibitem{sev}  W.J.L. Buyers, R.M. Morra, R.L. Armstrong, M.J. Hogan, P.
Gerlach, and K. Hirakawa, Phys. Rev. Lett. {\bf 56}, 371 (1986); J.P.
Renard, M. Verdaguer, L.P. Regnault, W.A.C. Erkelens, J. Rossat-Mignod, and
W.G. Stirling, Europhys. Lett. {\bf 3}, 945 (1987).; K. Katsumata, H. Hori,
T. Takeachi, M. Date, A. Yamagishi, and J.P. Renard, Phys. Rev. Lett. {\bf 63%
}, 86 (1989).

\bibitem{aklt}  I. Affleck, T. Kennedy, E.H. Lieb, and H. Tasaki, Phys. Rev.
Lett. {\bf 59}, 799 (1987); Commun Math Phys {\bf 115}, 477 (1988).

\bibitem{ni}  M. den Nijs and K. Rommelse, Phys. Rev. B {\bf 40}, 4709
(1989).

\bibitem{ken}  T. Kennedy and H. Tasaki, Phys. Rev. B {\bf 45}, 304 (1992).

\bibitem{gom}  G. G\'{o}mez-Santos, Phys. Rev. Lett. {\bf 63}, 790 (1989).

\bibitem{xu}  G. Xu, J.F. DiTusa, T. Ito, K.Oka, H. Takagi, C. Broholm, and
G. Aeppli, Phys. Rev. B {\bf 54, }R6827 (1996).

\bibitem{ram}  {\ }A.P. Ramirez, S.W. Cheong, and M.L. Kaplan, Phys. Rev.
Lett.{\bf 72}, 3108 (1994).

\bibitem{dmrg}  C.D. Batista, K. Hallberg, and A.A. Aligia, Phys. Rev. B 
{\bf 58 }, 9248 (1998). .

\bibitem{dit}  J.F. Di Tusa, S.-W. Cheong, J.-H. Park, G. Aeppli, C.
Broholm, and C.T. Chen, Phys. Rev. Lett. {\bf 73}, 1857 (1994).

\bibitem{koj}  K. Kojima, A. Keren, L.P. Le, G.M. Luke, B. Nachumi, W.D. Wu,
Y.J. Uemura, K. Kiyono, S. Miyasaka, H. Takagi, and S. Uchida, Phys. Rev.
Lett. {\bf 74}, 3471 (1995).

\bibitem{lu}  Z.Y. Lu, Z.B. Su, and Lu Yu, Phys. Rev. Lett. {\bf 74}, 4297
(1995).

\bibitem{sor}  E.S. Sorensen and I. Affleck, Phys. Rev. B {\bf 51, }16115
(1995).

\bibitem{pen}  K. Penc and H. Shiba, Phys. Rev. B {\bf 52 }, R715 (1995).

\bibitem{dag}  E. Dagotto, J. Riera, A. Sandvik, and A. Moreo, Phys. Rev.
Lett. {\bf 76}, 1731 (1996); C.D. Batista, A.A. Aligia and J. Eroles, {\it %
ibid }{\bf 81}, 4027 (1998); E. Dagotto and J. Riera, {\it ibid }{\bf 81},
4028 (1998).

\bibitem{epl}  C.D. Batista, A.A. Aligia and J. Eroles, Europhys. Lett. {\bf %
43}, 71 (1998).

\bibitem{vai}  A.A. Aligia, E.R. Gagliano, and P. Vairus, Phys. Rev. B {\bf %
52}, 13601 (1995).

\bibitem{whi}  S.R. White and D.A. Huse, Phys. Rev. B {\bf 48 }, 3844
(1993); T. Jolicoeur and R. Lacaze, Phys. Rev. B {\bf 50 }, 3037 (1994).

\bibitem{gol}  O. Golinelli, Th. Jolicoer, and R. Lacaze, Phys. Rev. B {\bf %
50}, 3037 (1994).

\bibitem{yan}  T. Yanagisawa and Y. Shimoi, Phys. Rev. Lett. {\bf 74}, 4939
(1995).

\bibitem{review}  H. Tsunetsugu, M. Sigrist and K. Ueda, Rev. Mod. Phys. 
{\bf 69}, 809 (1997).

\bibitem{penc}  K. Penc and R. Lacaze, cond-mat/9806116.

\bibitem{cie}  M. Cieplak, Phys. Rev. B {\bf 18 }, 3470 (1978); E.
M\"{u}ller-Hartmann and E. Dagotto, Phys. Rev. B {\bf 54}, R6819 (1996).

\bibitem{zan}  J. Zang, H. R\"{o}der, A.R. Bishop, and S.A. Trugman, J.
Phys.: Condens. Matter {\bf 9}, L157 (1997); references therein.

\bibitem{ino}  J. Inowe and S. Maekawa, Phys. Rev. Lett. {\bf 74}, 3407
(1995).

\bibitem{rie}  J. Riera, K. Hallberg, and E. Dagotto, Phys. Rev. Lett. {\bf %
79}, 713 (1997).

\bibitem{tsu}  H. Tsunetsugu, Y. Hatsugai and K. Ueda, Phys. Rev. B {\bf 46}%
, 3175 (1992).

\bibitem{fuji}  S. Fujimoto and N. Kawakami, J. Phys. Soc. Japan {\bf 66},
2157 (1997).

\bibitem{sug}  see for example S. Sugano, Y. Tanabe, H. Kamimura, {\it %
Multiplets of Transition-Metal Ions in Crystals }(volume 33 in {\it Pure and
Applied Physics}, Academic Press, New York and London, 1970).

\bibitem{h3b}  P.B. Littlewood, C.M. Varma and E. Abrahams, Phys. Rev. Lett. 
{\bf 63}, 2602 (1989); references therein; A.A. Aligia, Phys. Rev. B {\bf 39}%
, 6700 (1989).

\bibitem{con}  E.E. Condon and G.H. Shortley, {\it Theory of Atomic Spectra}%
, (Cambridge University Press, Cambridge and New York, 1935).

\bibitem{elp}  J. van Elp, H. Eskes, P. Kuiper, and G. A. Sawatzky, Phys.
Rev. B {\bf 45}, 1612 (1992).

\bibitem{hyb}  M.S. Hybertsen {\it et al.}, Phys. Rev. B {\bf 45 }, 10032
(1992); J.B. Grant and A.K. Mc Mahan, Phys. Rev. Lett. {\bf 66}, 488 (1991).

\bibitem{cell}  H.B. Sch\"{u}ttler and A.J. Fedro, Phys. Rev. B {\bf 45 }%
,7588 (1992); M. E. Simon, M. Bali\~{n}a and A. A. Aligia, Physica C {\bf 206%
}, 297 (1993); L.F. Feiner, J.H. Jefferson and R. Raimondi, Phys. Rev. B 
{\bf 53, } 8751 (1996); V.I. Belinicher, A.L. Chernyshev, and V.A. Shubin,
Phys. Rev. B {\bf 54}, 14914 (1996); M.E. Simon, A.A. Aligia and E.R.
Gagliano, Phys. Rev. B {\bf 56 }, 5637 (1997); references therein.

\bibitem{note}  This change of basis between the O $p_{z}$ orbitals lying
between Ni sites $i$ and $i+1$ ($p_{i+1/2,\sigma }$) and the O Wannier
functions $\pi _{j,\sigma }$ centered at Ni site $j$ is: $p_{i+1/2,\sigma
}=2\sum_{j}(-1)^{j+1}[\pi (2j-1)]^{-1}$ $\pi _{i+j,\sigma }$.

\bibitem{note3}  If some or all of the four nearest O atoms to the
transition-metal ion along the $\pm x$ and $\pm y$ directions are lacking,
the low-energy reduction procedure and the form of $H_{KH}$ (Eq. (\ref{e4}))
are not substantially modified. However, if one of the mirror symmetries
through the planes $x=0$ or $y=0$ is severely altered, the hopping of the
``localized '' $b_{i\sigma }$ holes, can no longer be neglected in $H_{KH}$.

\bibitem{note2}  For example, a large contribution to the hopping between
two nearest-neighbor cells with 2 and 3 holes respectively, is due to $\pi
_{j,\sigma }-\pi _{j+1,\sigma }$ hopping which is favored if both cells have
parallel total spin (antiparallel to that of the $\pi _{j,\sigma }$ hole $%
\sigma $). The amount of $\pi $ holes is much smaller in cells with 1 and 2
holes than in those with 3 holes.

\bibitem{tsun}  H. Tsunetsugu, Y. Hatsugai, K. Ueda, and M. Sigrist, Phys.
Rev. B {\bf 46}, 3175 (1992).
\end{references}
\end{document}